# Theory of Asymptotic Operation

## A summary of basic principles

Fyodor V. Tkachov

Institute for Nuclear Research of Russian Academy of Sciences
60-th October Ave. 7a, Moscow 117312, RUSSIA

This summary of several talks given in 1990–1993 discusses the problem of asymptotic expansions of multiloop Feynman diagrams in masses and external momenta — a central problem in perturbative quantum field theory. Basic principles of the theory of asymptotic operation — the most powerful tool for that purpose — are discussed. Its connection with the conventional methods is explained (the BPHZ theory, the method of leading logarithmic approximation etc.). The problem of non-Euclidean asymptotic regimes is discussed as well as ways of its solution.

## Introduction 1

The main purpose of this talk is to explain what asymptotic operation (=*As*-operation) is and how it is used in applications where higher-order corrections are studied and computed within the framework of perturbative QFT.

First of all, the place of the theory of *As*-operation in the elementary particle physics can be explained using the following analogy:

| Classical physics | High-energy physics |
|---|---|
| Newton's mechanics | Perturbative QFT |
| Theory of Hamiltonian systems | Theory of Feynman diagrams |
| Asymptotic methods of classical mechanics | **Theory of asymptotic operation** |
| Theory of differential equations | Theory of distributions |

1.1

In other words, the theory of *As*-operation considers analytical/calculational aspects of applied (perturbative) Quantum Field Theory which represents the formalism of the present day high-energy physics. On the other hand, (part of) the theory of *As*-operation can be regarded as a new applied branch of the theory of distributions.

Because all amplitudes are expressed in terms of Feynman diagrams in perturbative QFT in a more or less explicit fashion, the situation here is somewhat simpler than in classical mechanics. Correspondingly, the theory of *As*-operation is neither as difficult nor as vast as the theory of Hamiltonian systems. But it is difficult enough — and important enough.

I will try to answer the following questions:

- What is the problem?
- What is the asymptotic operation?
- What should be done to extend the theory of *As*-operation to the non-Euclidean case?

## What is the problem? 1.2

The answer to this question is going to be rather lengthy because the time spent on understanding the problem is the time well spent; there are enough examples of how detrimental can be getting down to business — i.e. to calculations, proofs, writing papers, etc. — too soon. In fact, the reanalysis from a novel point of view of the venerable and subtle problem of asymptotic expansions of Feynman diagrams in masses and momenta, is at the heart of the theory of *As*-operation.

♦ **Feynman diagrams** remain the best source of quantitative dynamical information in high-energy physics and this is hardly going to change soon (see proceedings of any workshop on physics programme of supercolliders).

♦ **Multiloop diagrams** have to be studied for the following reasons:
— some physical effects are entirely due to radiative (loop) corrections;
— taking into account higher-order corrections in QCD reduces the ambiguities due to truncation of perturbation series [97; 108];
— precision achieved in the studies of Standard Model is so high that computation of two-loop corrections with several different masses is required.

♦ **One measure of analytical complexity** of Feynman diagrams is the number of loop integrals *vs.* the number of external independent dimensionless parameters:

| *no. of loops:* | *no. of parameters one can handle:* |
|---|---|
| tree | any cross-section |
| 1 loop | many |
| 2 loops | a few (one) |
| 3 (4 RG) loops | zero (a single dim'l parameter) |
| 4 (5 RG) loops (**record**) | zero |
| more | hopeless (?!) |

1.3

A calculation is of (effective) $l$-loop complexity if it requires evaluation of finite parts of the corresponding diagrams. The effective number of loops may differ from the formal number of loops; thus, calculations of $\beta$- and other renormalization group functions require only calculation of divergent parts which is effectively equivalent to a reduction of the loop number by one [27]. On the other hand, in non-Euclidean problems the number of loops correspond to twice as many poles/large logarithms as in the Euclidean case because the time-like and space-like components play a different role. This should be taken into account when comparing Euclidean and non-Euclidean problems.

At the tree level, any cross-section can be calculated as a rational function of kinematic parameters. Calculations of this kind can be highly sophisticated and involve e.g. one-loop calculations to cancel infrared divergences. At the one-loop level, one can handle several problems with many parameters even if answers in the most general case may be very cumbersome [28; 75; 76]. At the two-loop level, only a few problems with one or at most two parameters are tractable (cf. calculations of the Leiden group [81]). At the three-loop level, only problems with no dimensionless parameters seem to be systematically tractable by analytical methods. One example is the $g_\mathrm{e}-2$ problem in QED [119, and refs. therein]. Another is massless propagator-type integrals that are systematically calculable through three loops [35; 43; 54]. There are two types of problems that can be reduced to such integrals: evaluation of coefficient functions of Wilson short-distance OPE ([45; 50]; the main application is to deeply inelastic lepton-nucleon scattering; e.g. [89]), and evaluation of four-loop renormalization group functions (see e.g. [27]). The effective reduction of the number of loops is due to a trick [16; 24; 27; 34] which involves retaining leading terms of an expansion in momenta and masses of the relevant diagrams — see below). A third class of problems related to the two already mentioned involves massive integrals [107; 115; 117]. Lastly, there are two examples of calculations of four-loop complexity: the formidable numerical calculation of $g_e - 2$ in QED [37; 47; 113; 114] which needs to be improved yet, precision-wise; the five-loop renormalization group calculation of $\beta$-function and anomalous dimensions in the scalar $\varphi^4_{D=4}$ model [51]. The latter calculation was done by hook or by crook — using various analytical tricks as well as a high-precision numerical summation in one of the diagrams (which was later reproduced analytically [48]); it has an interesting physical application to the theory of phase transitions (evaluation of the critical exponents via the $\varepsilon$-expansion method of Wilson and Fischer; cf. [30]). It is interesting that a somewhat slow convergence of the resummed series for one of the exponents ($\eta$; see [57]) warrants a six-loop calculation of the wave function anomalous dimension. Such a calculation may be feasible using a combination of analytical and numerical methods (there are not so many diagrams and most of them are trivial) but is still very hard.

♦ It should be clear that **reduction of the number of independent parameters** of the problem is the only way to go to higher orders of perturbation series. This can be achieved by exploiting large/small parameters (sometimes in a non-obvious way — I will present examples below) and performing asymptotic expansions with respect to them. This replaces a complicated initial function with a series in powers and logarithms of the expansion parameter(s). The coefficients of such a series are simpler functions (with lesser number of independent parameters) than the original one, and calculational advantages are often great — even if the number of such simpler functions can be large: "bad" complexity is thus traded for "trivial" complexity of a large volume of algebra.

♦ Finally, the problem of asymptotic expansions of Feynman diagrams has **two logical levels**. This is because physical quantities are expressed as sums over infinite collections of diagrams which (collections) have a hierarchical structure. The latter is manifest in the existence of structural equations (e.g. the renormalization group equations) that allow one to resum the corrections (see e.g. [58]). Under certain conditions asymptotic expansions inherit such a hierarchical structure so that it becomes possible to resum corrections using some analog of renormalization group equation (e.g. Altarelli-Parisi [23] or Balitsky-Kuraev-Lipatov-Fadin etc. [19; 21; 25]). Deriving such equations requires knowledge of the global structure of expansions for the entire perturbation series.

— At the **lower level**, one deals with expansions of individual diagrams. The problem here is **analytical** in nature.
— At the **higher level**, one deals with globally defined (i.e. defined in terms of entire collections of Feynman diagrams) Green functions. Their expansions should be recast into a global form (cf. OPE) that would allow a non-perturbative interpretation. Transition from the level of individual diagrams to the global level[i] is of a **combinatorial** nature and is not difficult *provided a solution of the right kind was found at the lower level*.

Solutions of the "right kind" are those which, first, exhibit the property of perfect factorization (see [45; 44; 111] and the discussion below). Second, expansions of diagrams should consist of such "pieces" of initial diagrams that would again have the form of multiloop diagrams so that one could — after a combinatorial regrouping of the "pieces" coming from different initial diagrams — obtain objects like matrix elements of some composite operators. It is this aspect that complicates the asymptotic expansion problem in perturbative QFT as compared to the general mathematical problem of expansion of integrals in parameters.

It is remarkable that the solutions of the "right kind" are only naturally described in the language of distributions[ii] (see the discussion below) whereas the older approaches stay strictly

---

[i] known as factorization or exponentiation; the term *globalization* would be less misleading.

[ii] There is a direct analogy with the problem of classical geometry about measuring the unit square's diagonal; its solution is also expressed in terms of "generalized" objects — irrational numbers. Note that discovery of such generalized solutions is invariably accompanied by confusion and all sorts of negative emotions among the

within the usual integral calculus. This explains why the older approaches encountered such difficulties with the expansion problem in QFT for general (non-Euclidean) asymptotic regimes.[i]

To summarize:

| | |
|---|---|
| **Evaluation of asymptotic expansions of multiloop Feynman diagrams** in masses and external momenta — expansions satisfying certain conditions (*perfect factorization*) — is **a central mathematical problem of perturbative QFT**. | 1.4 |

This problem in its entirety remains unsolved despite efforts of scores of theorists over 40 years and many partial results obtained. The pathos of this text is that the theory of *As*-operation is currently the best hope to make a substantial progress towards a complete solution of the problem in a near future.

Before we discuss what such a hope is based upon, I would like to present a few examples of specific physical problems that involve asymptotic expansions of Feynman diagrams. This seems necessary because, on the one hand, mathematical physicists tend to limit the scope of their research to a few "canonical" problems like proving that expansions of diagrams run in powers and logs of the expansion parameter or short-distance operator product expansion[ii] — the problems where "rigorous" treatment is possible.

Unfortunately, (formal) rigor per se is of little significance outside student papers, and given a good analytical idea, formal rigor can always be added by any competent professional [3]. Moreover, applications are, as a rule, so cumbersome that obtaining fully formalized proofs that were actually accessible to the theoretical community for examination, remains a wishful thinking. Then the term "rigor" acquires a mythological aspect — and often used as a means of self-differentiation — being a sign of absence of genuinely new ideas. What is really needed is *correct* results — irrespective of whether accompanied by completely formal proofs or not — as well as formalisms that allow one to obtain such results. Correctness is not guaranteed by formal "rigor" but by precise understanding of the structures encountered in a problem — an understanding adequately reflected in the methods used to solve it.

On the other hand, there are limits to the liberties one is allowed to take with formulas: one's manipulations of formulas must remain mathematically meaningful to the extent warranted by the problems one deals with. The context I will have in view is the calculational problems of perturbative QFT of the advanced kind (typically, involving precise computations of higher-order corrections). In such a context, one may, e.g. — after exhibiting a recursive pattern, — satisfy oneself with a few typical examples to see how the induction continues rather than devising a complex notation for tackling the general case.[iii] But it is completely unacceptable e.g. not to clarify such issues as what, exactly, is the asymptotic re-

---

adherents of the old methods. The theory of asymptotic operation is no exception in this respect.

[i] However, since any expression with distributions can be rewritten in terms of ordinary integrals, the old methods could in principle be used for verification of results — *provided the answers have been found by another method***.** But such a verification is only possible in simple cases because it is extremely cumbersome. Even in the case of Euclidean asymptotic regimes it proved practically impossible to perform it explicitly (see the discussion below). Anyhow, practical value of such verifications is negligible (cf. the discussion in [110]).

[ii] Incidentally, phenomenologically meaningful problems are always formulated in the language of momentum representation, so that a literal acceptance of phrases like "short distances" etc. without a concrete analysis of the calculational reality only lead to the proving of useless — however rigorous — theorems.

[iii] This difference can be illustrated by the papers [101; 102] (a completely rigorous and formal regularization-independent theory of Euclidean asymptotic operation) and [111] (a derivation that is not formally rigorous but fully explicit, and relying on dimensional regularization to avoid syntactic complexities); a look at the formulas and reasoning in the two cases should give one a good idea of what I have in view. On the other hand, often enough the "proofs" of "factorization theorems" are such that one cannot help wondering whether their authors are at all familiar with the concept of manipulating formulae via a well-defined transformations rather than via a lot of pictures and pseudophysical handwaving.

gime one is dealing with, what are the small parameters of the problem, etc. Unfortunately, omissions of this sort are rather common in phenomenologically oriented papers.

## Examples of physical problems 2

First of all, asymptotic regimes can be divided into Euclidean and non-Euclidean. Euclidean regimes are those where the specifics of the problem allow one to perform Wick rotation without crossing causal poles and deal with purely Euclidean diagrams. The class of Euclidean regimes includes two typical cases: the familiar Wilson short-distance regime, and the heavy-mass regime where one considers low-energy amplitudes involving only light fields (e.g. effective low-energy weak interaction Lagrangians of light quarks [18; 17]). There are also mixed cases. An exhaustive phenomenologically meaningful solution of the asymptotic expansion problem for general Euclidean regimes was first provided by the theory of *As*-operation ([111] and refs. therein). The phrase “phenomenologically meaningful” translates into the technical requirement of *perfect factorization* that the expansions must satisfy. I will discuss the requirement later on.

### Deeply inelastic scattering 2.1

… of leptons off nucleons is well-known (Fig. 2.2; see e.g. the review [79]). The cross-section is parameterized by form-factors that are functions $F(x,Q^2)$ of two kinematic variables: the dimensionless Bjorken variable $x$ and the transferred momentum $Q^2=-q^2$. The standard asymptotic regime corresponds to $Q^2\to+\infty$ with fixed $x$. The asymptotic behavior is described by

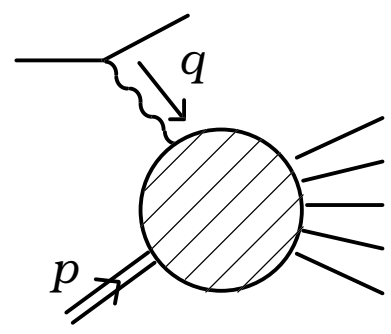

2.2

$$F(x,Q^2)\underset{\substack{Q^2\to+\infty\\ x\ \text{fixed}}}{\to} F_{\text{GLAP}}(x,Q^2)\times(1+O(Q^{-2}))\,, \qquad 2.3$$

where the leading term depends on $Q^2$ logarithmically, and the function $F_{\text{GLAP}}$ satisfies an integro-differential equation known as the Gribov-Lipatov-Altarelli-Parisi equation. Power corrections (i.e. corrections of order $O(Q^{-2})$; they are also known as higher-twist corrections) satisfy similar equations. This is a rare case where the structure of higher-twist corrections is known; this is because this problem is directly connected with the OPE at short distances.

The corresponding asymptotic regime can be described by saying that $p$ goes to a light-like value with $q$ fixed at a space-like value:

$$p\to\tilde{p},\quad \tilde{p}^2=0,\quad q\ \text{fixed},\quad q^2<0,\quad m_i=O(p-\tilde{p})\,. \qquad 2.4$$

The masses vanish at the same rate as $p$ approaches its light-like limiting value. (In the coordinate representation this corresponds to the light-cone limit [49].)

Note how $p$ tends to a vector that is light-like but non-zero component-wise. This is a characteristic feature of non-Euclidean asymptotic regimes.

Alternatively, one can consider moments of structure functions:

$$\int_0^1 \mathrm{d}x\, x^n F(x,Q^2)\underset{\substack{Q^2\to+\infty\\ n\ \text{fixed}}}{\to} C_n(Q^2)\times M_n\times(1+O(Q^{-2}))\,. \qquad 2.5$$

This asymptotic regime is essentially Euclidean. The coefficient functions $C_n(Q^2)$ are Fourier transforms of the coefficients of the Wilson OPE.

The two asymptotic regimes are not equivalent. This is reflected in the fact that reconstruction of structure functions from their moments is an ill-posed problem. This means that reconstruction of structure functions from their moments is only possible either if *all* moments are known exactly (as in reconstruction of the Altarelli-Parisi kernels from OPE anomalous dimensions), or if additional information is used in conjunction with numerical values of the first few moments.

### Small-$x$ problem and the Regge limit 2.6

(See e.g. the review [109].) One considers structure functions of deeply inelastic scattering at small $x$ where the amplitude is expected to be most sensitive to self-interaction of gluons:

$$F(x,Q^2) \quad \text{for } x\to 0 \text{ with } Q^2 \text{ fixed}. \tag{2.7}$$

This is equivalent to studying the Regge limit:

$$s=(p+q)^2 \gg Q^2,\, p^2,\, m_i^2 . \tag{2.8}$$

Via the optical theorem, this is connected with the amplitude of an elastic process:

$$F(s,t,u,m_i^2) \underset{s\to+\infty}{\approx} f(s,t,u,m_i^2)\times(1+O(s^{-1})) , \tag{2.9}$$

where the function $f$'s dependence on $s$ is logarithmic. The problem here is to determine the analytic form of the dependence on $s$, i.e. the form of $f(s,\dots)$.

The Regge limit was extensively but inconclusively studied in connection with the theory of dispersion relations and various bounds on asymptotic behavior of cross-sections since the '50s. The theoretical result available within pQFT is known as the BFKL equation [20; 21; 25]. The equation is derived using the so-called technique of leading logarithm approximation [2; 74]. Moreover, such an equation is analogous to the renormalization-group equation for coefficient functions of short-distance OPE in the one-loop approximation, and nobody knows how it has to be modified beyond the leading logarithm approximation. The question is of great practical significance because the leading logarithm approximation is sometimes a poor one unless one goes to energies much higher than one can afford. [i] It should also be mentioned that there is a considerable body of expertise and interesting physical results obtained by the leading-logarithm sect that remains, to a great extent, shrouded in mystery to outsiders.

### QCD-improved parton model 2.10

This is a collection of prescriptions that describes the structure of leading-twist terms in asymptotic expansions of amplitudes of a class of (semi-) inclusive processes for certain asymptotic regimes (see e.g. [46]). The underpinning theoretical results are known as “factorization theorems" [65]. Their status is far from being satisfactory because of the gaps in their derivation. Thus, for instance, proving smallness of the remainder of the expansion is technically very similar to evaluating the next correction, whereas nothing is known about power corrections in problems like Sudakov formfactors. What has been lacking so far is an exact formalism to deal with both analytical and combinatorial aspects of the problem. The theory of Euclidean *As*-operation provides a model of what such a formalism might look like.

For example, within QCD following [31], consider a process $H_1+H_2\to H_3+X$ where $X$ is an inclusive state. There are three independent kinematic parameters in the problem: the Mandelstam variables $S$, $T$ and the invariant mass of the inclusive state $M_X^2$. The variable $U$ is determined from the relation

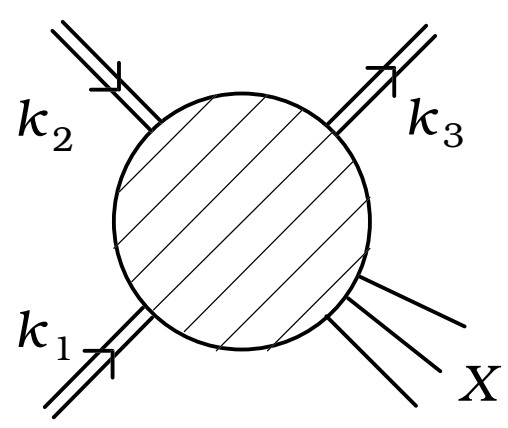

2.11

$$S+T+U = M_X^2 + m_1^2 + m_2^2 + m_3^2 , \tag{2.12}$$

---

[i] I thank L. Frankfurt for a discussion of this point.

where $m_i^2$ are the masses of the three particles. There are also small parameters like quark masses that are not indicated. One normally considers such processes in the regime $S, T, M_X^2 >> m_i^2$ which corresponds to the standard QCD-improved parton model. Then the cross-section is represented as a convolution of a cross-section of a "hard parton subprocess" with parton distributions that are similar to the distributions $F(x,Q^2)$ that emerged in the study of deeply inelastic scattering and satisfy the same integro-differential equations.

The essential dynamical information is concentrated in the hard cross-section which is the analog of the OPE coefficient functions. In the present case it depends on two dimensionless parameters — $S$ is normally scaled out of all expressions.

## Example of a hidden small parameter 2.13

Calculations of the hard cross-section of the preceding example are rather cumbersome already at the one-loop level [31]. However, one can also consider the quasi-elastic asymptotic regime $S, T >> m_i^2, M_X^2$, then one has to deal only with one independent dimensionless parameter. (In practice one deals with a version corresponding to $S, T >> M_X^2 >> m_i^2$ .) This is motivated by the following fact. As has been mentioned above, the parton-model formula for the cross-section involves a convolution of the hard cross-section and the parton distributions $F_i(x,Q^2)$ that are power-suppressed as $(1-x)^n$ near $x=1$. It turns out that the region where such a suppression is not effective corresponds to the quasi-elastic region; moreover, the contributions that are dominant in the quasi-elastic region are enhanced by singularities (see e.g. [106] and refs. therein). It turns out that calculations of one-loop corrections then become incomparably simpler while the corresponding approximations based on retaining only leading terms work surprisingly well (within numerical accuracy 5%) all the way down to $x_T \gtrsim 0.2$ [106]. This means that the problem contains an effective hidden small parameter $M_X^2/S$. Note that the "small" parameter need not be numerically small; it is only important that one performs an expansion with respect to such a parameter, and that the expansion works within desired accuracy for the values of interest.

There is no general theory of the quasi-elastic regime; only some leading-logarithm results are available (see e.g. [95]). Note that in the small-$x$ region the parton model is not expected to work anyway. On the other hand, the quasi-elastic regime is, in principle, no worse than the standard parton-model regime.

## Sudakov formfactor 2.14

In the theory of Sudakov formfactor [2] one studies the behavior of formfactors of electrons, quarks and other elementary fields at large momentum transfers (Fig. 2.15). Results for the leading power term are described in [64]. An interesting fact is that a new type of operators occur here (string operators [64]) whereas only local operators occur in the Wilson expansion an short distances. String operators also occur in other problems [104].[i]

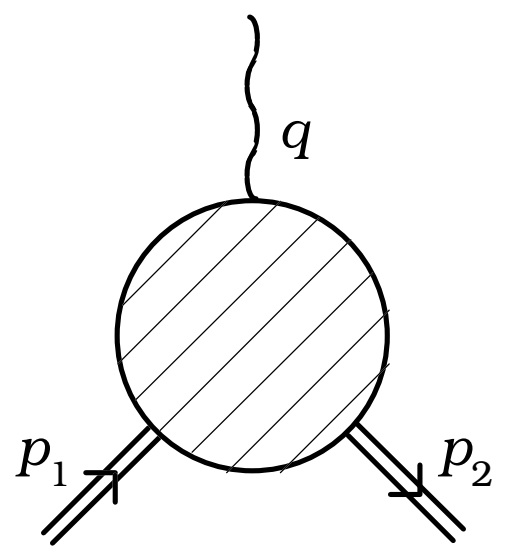

2.15

Note that the Sudakov formfactor decreases at large $Q^2$ — the effect known as Sudakov suppression. The importance of the Sudakov formfactor problem is due to an old hypothesis that the Sudakov suppression may play for some asymptotic regimes a role similar to asymptotic freedom (cf. [96; 100]). This problem remains unclarified.

---

[i] I thank A.V. Radyushkin for a discussion of this point.

The asymptotic regime here can be described as follows (cf. 2.4):

$$p_{1,2} \to \tilde{p}_{1,2}, \quad \text{where} \quad \tilde{p}_{1,2}^2 = 0, \quad \text{at a fixed } q \text{ and } m = O(p - \tilde{p}) \qquad 2.16$$

## QCD jets 2.17

The concept of jets in the processes at high energies is the key element of the important paradigm in the physics of elementary particles that has given rise to many interesting problems (see e.g. the review [116]). In practice, one deals here with cross sections of parton type with many soft and collinear singularities.

Below an example of asymptotic expansion problem is given, which is rather different from those already considered.

## The double-box problem[i] 2.18

Consider a process with three hadron jets in the final state. To order $O(\overline{\alpha}_s^3)$ there are 3-parton contributions of 2-loop complexity. Ones that are particularly hard to compute analytically are the so-called double-box diagrams and their non-planar analogues (Fig. 2.19; the nature of the virtual lines — gluons, quarks etc. — is not important as long as they are massless). Such diagrams are functions of $M^2 > 0$ and the invariant masses of the three final state partons, $s_{ij} = (k_i + k_j)^2$, $\{i,j\} = \{1,2\},\{1,3\},\{2,3\}$. However, since $s_{12} + s_{13} + s_{23} = M^2$, there are only two independent massless parameters in the problem, e.g.

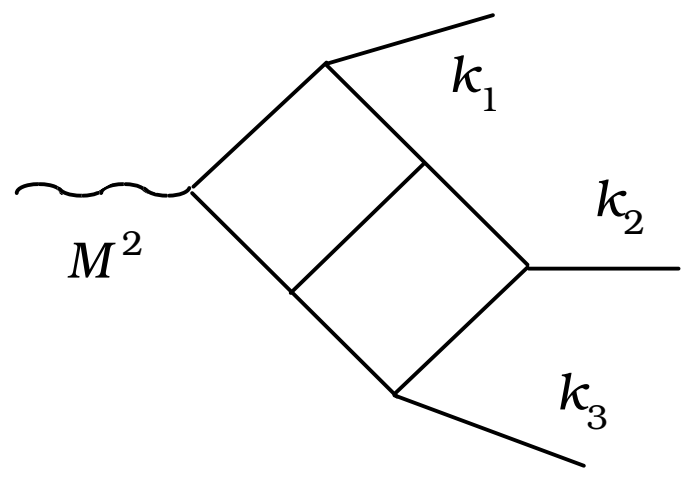

2.19

$$x = s_{12}/M^2 > 0, \; y = s_{23}/M^2 > 0, \; x + y < 1. \qquad 2.20$$

Such an amplitude has IR singularities (the maximal singularity $\varepsilon^{-4}$). A gauge-invariant cancellation of all IR singularities in the final answer can be achieved by using the dimensional regularization [11]. Another requirement is that the result will be used in a Monte Carlo event generator to produce something like $O(10^5)$ events, so that the answer should represent an efficient computational algorithm.

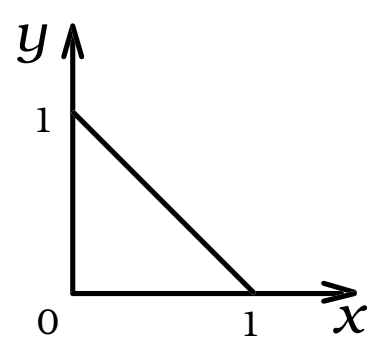

2.21

A possible scenario for a systematic treatment of such diagrams — a scenario that is applicable in many other cases — consists in representing them as expansions in powers and logarithms of $x, y, 1-x, 1-y$.

To clarify this point, note the following. It is often incorrectly assumed that "exact" expressions for the integrals being studied in terms of special functions (polylogarithms etc.) are the final answer which one should aim at in an "analytical" calculation. But a somewhat more "final" answer would consist of asymptotic formulas at singular points (including points at infinity) plus efficient numerical algorithms for the regular points, whereas "explicit" representations in terms of special functions are only useful inasmuch as they can facilitate computations. One example are the expressions in terms of infinite multifold series in generalized hypergeometric functions [105]. Another example is the numerical calculations described in [103] where computations of dilogarithms to 15 digits were sometimes required in order to obtain numerical answers with a reasonable accuracy. However, if there is a direct and systematic method for obtaining asymptotic expansions near singular points, then in combination with numerical algorithms (using e.g. Feynman parameters) this may allow one to obtain all the necessary information bypassing the "exact analytical" calculation. [Note added: the "algorithm of algorithms" of algebraic multiloop calculations found in [118] allows one to express any Feynman-parameterized integral in an absolutely convergent form with all the poles explicitly extracted.]

---

[i] I thank R.K. Ellis and W.T. Giele for explaining to me the double-box problem.

In the present case, say, the expansion in $x$ results in a series whose coefficients are calculable functions of $y$. To obtain such an expansion, one needs a non-Euclidean extension of the technique of asymptotic operation. One can also seek expansions for asymptotic regimes $0 < x \ll y \ll 1$ etc. That way one could obtain representations of the amplitude in the form of expansions in $x$ and $y$ near the boundaries of the kinematic triangle. Convergence of such expansions is, roughly, geometric. If the convergence is not sufficiently fast near the center of the triangle, one can always construct a simple interpolation formula based on values of a few points, obtained numerically (the amplitude is analytic inside the triangle).

Measuring computational complexity by the number of poles in $\varepsilon = D - 4$, the described scenario corresponds to the record-setting 5-loop renormalization group (Euclidean) calculations of [51]. However, the non-Euclidean complications make the problem much more cumbersome, whereas the non-abelian nature of QCD causes a blowup of the algebra involved. Nevertheless, this is the only realistic scenario of tackling the double-box problem.

There is a remarkable recent analytical result [99] for the scalar double box but only for the planar case. Moreover, the answer of [99] is only for $D = 4$ and all $k_i^2 > 0$ to regulate the singularities. To translate the 4-dimensional results with non-zero $k_i^2$ to the required form (dimensional regularization with $k_i^2 = 0$), one could perform an expansion for $k_i^2 \to 0$. (Note that taking off a regularization for singular products is similar to asymptotic expansion in the sense of distributions, which underscores the fundamental role of that notion; cf. a discussion below.) It does not seem likely that a combination of tricks used in the calculation of the planar diagram [99] could prove useful in the non-planar case (cf. [98]). Then one would have to recur to the complete scenario based on expansions described above.

Lastly, at some point one may need to study the singular boundary regions more closely. Then one would need global results like factorization theorems and the corresponding structural equations for such asymptotic regimes — which represents another variation of the asymptotic expansions theme.

There are many more physical problems involving asymptotic expansions of Feynman diagrams, which cannot be discussed here, e.g. the effective heavy-quark Lagrangians [92] etc.

## General formulation of the asymptotic expansion problem 2.22

First of all, one has to reformulate the specific problem in terms of masses and external momenta instead of scalar kinematic variables, and to identify the asymptotic regime, i.e. which masses and momentum components are considered as small compared with others. One writes $Q, M >> k, m$. It is more convenient for technical reasons to work with small parameters with respect to which to expand. Due to homogeneity of amplitudes, both ways to state the problem are equivalent.

| $Q, M >> k, m$ | |
|---|---|
| $Q,\ M$ fixed | $k, m \to 0$ |
| large parameters | small parameters |

2.23

It should be stressed that the expression $Q, M >> k, m$ that describes the asymptotic regime need not be valid in the naive numerical sense. It only means that one intends to replace the initial function by its expansion with respect to the ratio of the two scales. The radius of convergence — or, rather, of numerical usefulness — of the resulting expansion can be quite large (cf. the examples of specific problems above).

*UV renormalization* introduces an additional dimensional parameter, $\mu_{\text{ren}}$. Without much loss of generality one can assume that all amplitudes are renormalized using the MS prescription (or any other massless renormalization scheme). Then renormalized diagrams are polynomials of $\log \mu_{\text{ren}}$ and it is inessential whether $\mu_{\text{ren}}$ is considered as small or large. I will ignore $\mu_{\text{ren}}$ in what follows.

If $A(Q, M, k, m)$ is the amplitude (Green function) to be studied then one wishes to obtain an expansion of the form:

$$A(Q, M, k, m) \cong \sum_i C_i(Q, M, \mu_{\text{fact}}) D_i(k, m, \mu_{\text{fact}}) \ , \qquad 2.24$$

where $C_i$ are conventionally referred to as "coefficient functions" while $D_i$, as "matrix elements" — the terminology inherited from Wilson's OPE.

I have explicitly indicated the presence of another parameter, $\mu_{\text{fact}}$, that emerges in the process of factorization of large and small parameters (typically: $\log(Q^2/m^2) \to \log(Q^2/\mu_{\text{fact}}^2) + \log(\mu_{\text{fact}}^2/m^2)$). It may or may not be set equal to the renormalization parameter.

## Perfect factorization 2.25

An extremely important requirement that was completely overlooked in the conventional understanding of the expansion problem is that of *perfect factorization* of large and small parameters [45]. It can be explained as follows. First, $C_i$ should contain only contributions proportional to the same power of the large scale. Second, it should not depend on the small parameters, *k* and *m*. Due to homogeneity of *A*, similar statements — *mutatis mutandis* — would be true for $D_i$. As an implication, $\mu_{\text{fact}}$ can no longer be naively interpreted as a cutoff in position space separating regions of integration momenta because using such cutoffs generates terms with power dependence on $\mu_{\text{fact}}$. (A term like "lazy cutoff" due to A.V.Radyushkin would be more adequate.) The nature of $\mu_{\text{fact}}$ can be best described as follows. The expressions for $D_i$ turn out to be matrix elements of some complicated operators that are automatically supplied with an appropriate UV renormalization. If the requirement of perfect factorization is satisfied then it automatically turns out that such a renormalization uses one of the so-called massless schemes, e.g. the MS scheme and $\mu_{\text{fact}}$ is a renormalization parameter of such scheme (cf. the construction of *As*-operation in [91; 101; 102]).

At the level of individual diagrams, perfect factorization means that the expansions should be of a pure power-and-log type.

Because expansions of perturbative Green functions are obtained by a combinatorial regrouping of expansions of individual diagrams, the meaning of perfect factorization should be clarified at the level of individual diagrams. Here perfect factorization means that the expansions must run in pure powers and logarithms of the expansion parameter, and at first glance the requirement may seem meaningless. In fact, the analytical nature of multiloop diagrams is not that complex — they are integrals of rational functions. Calculational experience indicates that their asymptotic expansions in momenta and masses — e.g. extracted from "explicit" expressions via special functions or by splitting integration regions etc. — always run in powers and logarithms of the small parameter. Formal proofs of this fact for various asymptotic regimes were given in [14; 32; 36; 40].[i] However, the essence of the problem of expansions of multiloop diagrams consists in splitting the original integral — prior to explicit calculation, which represents a separate problem with its own specifics — into pieces so the some pieces contained the power-and-log dependence on the small expansion parameter and others, on the large ("spectator") parameters. The pieces should be objects that have a form of multiloop diagrams so that one could, after a combinatorial rearrangement of pieces coming from different unexpanded integrals, to rearrange them into objects like matrix elements of some composite operators. The papers cited above failed to construct such an expansion (such a problem was not even considered). On the other hand, in the papers [10; 12], where the expansions in operator form were constructed for the asymptotic regime of short distances, the "pieces" did not have purely power-and-log dependence on the small parameter.

---

[i] The analytical idea behind all such proofs is quite simple: multiplications by rational functions and one-dimensional integrations do not take one outside the class of functions with power-and-log asymptotics. The main difficulty is to carry out a more or less explicit reduction of the problem to sequences of one-dimensional integrations, and construct an argument for an arbitrary diagram from a sufficiently wide class. The cleanest way to organize such a reasoning (which also leads to substantially stronger results) is given by the theory of asymptotic operation and is based on the use of recursion in the problem (see below).

As an illustration, consider the following one-dimensional model of a one-loop integral that depends on a large "momentum" $Q$, a small "momentum" $k$, and a small non-zero mass $m$:

$$I(Q,k,m) = \int_0^\infty \mathrm{d}p \frac{1}{p+k+m} \times \frac{1}{p+Q+m} . \qquad 2.26$$

One has to expand the integral at $Q \to \infty$. There are three constructions possible.

1) The most straightforward approach consists in splitting the integration region as follows:

$$\int_0^{+\infty} \mathrm{d}p = \int_0^{\mu} \mathrm{d}p + \int_{\mu}^{+\infty} \mathrm{d}p , \qquad 2.27$$

and expand the integrand in $Q \to \infty$ in the first subregion, and expand in $k,m \to 0$ in the second subregion. All integrals here are calculable, and one can see that the resulting expansion will contain only powers and logs of the parameters, and there will also be powers of $\mu$ which cancel after adding contributions from the two subregions. The latter fact is indicative of something important being missed in this approach.

2) A standard solution of the kind obtained in [10] is as follows. First one performs an auxiliary subtraction:

$$I(Q,k,m) = \int_0^\infty \mathrm{d}p \left[ \frac{1}{p+k+m} - \frac{1}{p+m} \right] \times \frac{1}{p+Q+m} + \int_0^\infty \mathrm{d}p \frac{1}{p+m} \times \frac{1}{p+Q+m} . \qquad 2.28$$

Now in the first term, one can replace the propagator depending on $Q$, by an expression that corresponds to asymptotics at large $Q$, as follows:

$$I(Q,k,m) = \left\{ \frac{1}{Q+m} \right\} \int_0^\infty \mathrm{d}p \left[ \frac{1}{p+k+m} - \frac{1}{p+m} \right] + \left\{ \int_0^\infty \mathrm{d}p \frac{1}{p+m} \times \frac{1}{p+Q+m} \right\} + O(Q^{-2}) . \qquad 2.29$$

The expression in square brackets corresponds to a renormalized matrix element of a local operator (subtraction at zero $k$ according to the MOM scheme [58]). The curly braces contain the pieces that depend on $Q$ (which corresponds to "short distance" after Fourier transform) but not on $k$ ("large distances") — such pieces form coefficient functions eventually.

Although the problem as posed by Wilson [8] is formally solved — short and long distances are factorized, but phenomenological applications require more than just that, for the following reasons:

First, physical problems are formulated in momentum representation, and a correct asymptotic regime is $Q \gg k,m$, i.e. the mass should be a small parameter (unless explicitly speci-fied otherwise).

Second, an explicit calculation of the integrals shows that the asymptotics at $Q \to \infty$ of the "coefficient functions" contain higher power terms, i.e. ones suppressed by $O(m/Q)$ compared to the leading term. In asymptotically free theories coefficients of such terms are not calculable within perturbation theory [44].

Third, a rather complex dependence on $m$ in the coefficient functions makes calculations of perturbative corrections practically impossible.

Fourth, the method does not yield satisfactory results in theories with massless particles (e.g. QCD) — this is manifested in the fact that it is impossible to put $m = 0$ in 2.29 because of the non-integrable singularity at $p = 0$.

Lastly, for non-Euclidean regimes where singularities are localized on non-linear manifolds (light cone), the approach with presubtractions proves to be extremely inflexible. One needs, essentially, to know the answer beforehand in order to perform a presubtraction similar to the above. This circumstance — absence of sufficiently powerful heuristics for more complex problems — is the crucial one for appraising the standard approach.

3) Solutions that are devoid of the drawbacks discussed above were first obtained within the framework of the theory of asymptotic operation [45] (see also [111; 112] and refs. therein). To compare the answers obtained here with the expressions given above, only the final result is presented here, although the way of reasoning that automatically leads to such answers will be discussed in detail later on:

$$I(Q,k,m) = \left\{\frac{1}{Q}\right\}\int_0^\infty \mathrm{d}p \left[\frac{1}{p+k+m} - \left\{\frac{1}{p}\right\}^r\right] + \left\{\int_0^\infty \mathrm{d}p \left\{\frac{1}{p}\right\}^r \times \frac{1}{p+Q}\right\} + O(Q^{-2}) \,. \qquad 2.30$$

The expression $\{p^{-1}\}^r$ is a distribution defined according to the following formula:

$$\int_0^\infty \mathrm{d}p \left\{\frac{1}{p}\right\}^r \times \varphi(p) = \int_0^\infty \mathrm{d}p \frac{1}{p} \times [\varphi(p) - \theta(p/\mu_{\mathrm{fact}})\varphi(0)] \qquad 2.31$$

where $\theta(p/\mu)$ is a cutoff function equal to 1 at $p \lesssim \mu$ and to 0 at $p \geq \mu$. In the more general case there occur distributions with simple scaling properties (as in our example), which fact results in power-and-log dependence on $Q$ [101; 111]. In the above example, this can be verified by explicit calculation.

The subtraction in square brackets correspond to subtractions in a UV renormalization scheme of the MS type [91; 102]. If one uses an intermediate regularization (e.g. similar to the dimensional one) the following representations are valid:

$$\left\{\frac{1}{p}\right\}^r = \frac{1}{p} + \frac{\mathrm{const}}{\varepsilon}\delta(p) \,. \qquad 2.32$$

In other words, compared to the "bare" expansion obtained from the original integral by expansion in $m$, there occur "counterterms" localized at the points of singularities. If one uses such representations (general formulas for this case are presented in [102; 111; 112]), then in calculations it is sufficient to deal — apart from the simple terms with $\delta$-functions — with massless integrals. For the coefficient functions of Wilson's OPE one obtains massless self-energy integrals [45; 50], and for such integrals there exist an efficient algorithm that can handle 3-loop integrals [35; 43; 54; 72; 87].

Although in this simple example one can rewrite the result in such a way as to get rid of the expressions containing distributions, in more complex cases (with more integration momenta) this is harder to achieve, and also the heuristic connection with the derivation would be lost.

So one concludes that the expansions that possess the property of perfect factorization, first, can be obtained in a form suitable for performing the combinatorial factorization. Second, such expansions are described most naturally in the language of distributions.

Finally, the requirement of perfect factorization is of great importance, both technically and conceptually. This is due to the fact that such expansions are unique [111]. Indeed, such expansions inherit properties like gauge identities, unitarity[i] etc. from the initial amplitude in an orderly manner, *automatically*, and one does not have to worry about, say, that the final expressions should not contradict unitarity. *All one has to worry about is that the expansions run in pure powers and logarithms of the expansion parameter*. On the other hand, the power-and-log nature of the expansions is preserved in the recursive construction of the expansions typical of the theory of *As*-operation because the product of power-and-log series is again a series of power-and-log type.

Also, if an expansion does not satisfy the requirement of perfect factorization then coefficient functions contain higher-twist contributions which are not amenable to perturbative calculation (for the case of OPE this was observed in [44]). Recall that the very usefulness of OPE is rooted in perturbative calculability of coefficient functions owing to asymptotic freedom in QCD.

## Understanding the problem from mathematical point of view 3

...the perturbative version of the Wilson short distance expansion is completely clear since more than ten years. ...A new theory is not needed.

*A referee for Nuclear Physics B*

*"Do not understand me too quickly"*

*Chinese saying*

### Why distributions?

Consider the following integral that has to be expanded in $m \to 0$ (the infinitesimal imaginary terms in denominators are omitted for brevity):

$$\int \mathrm{d}p \frac{1}{(p^2 - m^2)(p-Q)^2}\,. \tag{3.1}$$

The well-known difficulty with expanding such an integral is that the *m*-dependent propagator develops non-integrable and progressively more severe singularities when formally expanded in *m*:

$$\frac{1}{p^2 - m^2} = \frac{1}{p^2} + \frac{m^2}{p^4} + O(m^4)\,. \tag{3.2}$$

The usual way to avoid this is to split integration domain so as to isolate the dangerous regions and/or perform appropriate subtractions at $p = 0$ of the factors that do not participate in the expansion in *m*, in order to neutralize those singularities. This is the idea behind the conventional approaches — either in Zimmermann's implementation [10; 12] or the techniques of "leading integration regions" [2; 74].

There is nothing wrong with such an approach *per se*. But in the multiloop case the additional subtractions proliferate and tend to get out of hand. This is why the forest formula [7; 71; 10] (which fi-nally provided a closed formal description of the resulting entanglement of subtractions for the simplest case of Euclidean expansions) came to be regarded as a model for the general non-Euclidean case. As I will explain shortly, it would be incorrect to interpret the prospective extension of the *As*-operation to the non-Euclidean case as a construction of non-Euclidean forest formula because the non-Euclidean forest formula may not even exist in an intelligible form.

[i] I am grateful to L.N. Lipatov for a discussion of this point.

Given that the non-Euclidean forest formula failed to materialize in more than twenty years, let us subject the problem to a scrutiny — the single most powerful problem-solving method.

First of all, difficult problems should not be considered in isolation, nor their important aspects ignored. Thus, one could recall that it makes little sense to consider an isolated diagram within applied QFT where one deals with infinite, recursively organized (sic!) collections of such diagrams. Even better, there are many models that give rise to similar integrals. For instance, consider the following variant of 3.1 which differs from 3.1 by a presence of a heavy mass:

$$\int \mathrm{d}p \frac{1}{(p^2 - m^2)((p-Q)^2 - M^2)} . \qquad 3.3$$

One can observe that the troublesome *m*-dependent factor is the same as in 3.1 while the "passive" factor that is smooth at the singular points of the first factor, is different. A natural guess would be that the exact form of the "passive" factor is irrelevant — as long as it is well behaved:

$$\int \mathrm{d}p \, \frac{1}{(p^2 - m^2)} \, \varphi(p) , \qquad 3.4$$

where $\varphi(p)$ is smooth everywhere and decreases at large $p$ fast enough to ensure convergence (in order to avoid collateral complications of UV convergence) [i]. It is clear that both the expression 3.4 and its expansion in $m$ — whatever the form of such expansion — are linear with respect to $\varphi(p)$. To expand 3.4 for well behaved $\varphi(p)$ is exactly the same as *to expand the propagator* $(p^2 - m^2)^{-1}$ *in the sense of distributions*.

Let me immediately explain what is the difference between expansions in the sense of distributions and formal Taylor expansions, from practical point of view. The key result here is the so-called *extension principle* [39; 111] — a constructive if abstract (and, basically, very simple) proposition modeled after the classic Hahn-Banach theorem on extension of functionals. The extension principle states that, given an *m*-dependent functional defined on a linear space, and another functional that approximates the first one but only on a subspace, then the approximating functional can be extended to the entire space with its approximation property preserved. Practically, the prescription of the extension principle consists, loosely speaking, in adding to the formal expansion counterterms localized at the singular points, therefore, proportional to $\delta$-functions and their derivatives. Similarly to the case of the *R*-operation in coordinate representation, the coefficients of the counterterms are divergent to counterbalance the non-integrable singularities of the formal expansion. In contrast with the *R*-operation, however, *the finite parts of the coefficients are not arbitrary* because they should be adjusted to ensure the approximation properties of the resulting expansion in the sense of distributions. In fact, if the expansion is to have the desired analytic form (in our case, the requirement of perfect factorization implies that they should contain only powers and logarithms of the expansion parameter) then the coefficients of the *counterterms are determined uniquely*. A practical recipe for determining the coefficients is given by the *consistency conditions* [39; 111].

In the present case, the *m*-dependent functional is the distribution corresponding to the non-expanded propagator 3.4; the space is the space of all test functions; the approximating functional is the formal

[i] At this point it should be emphasized that the factor in the original integral that was replaced with a test function, is strictly speaking not itself a test function: We are discussing here a heuristic that allows one to focus on the element of the problem.

Taylor expansion 3.2; the subspace on which the latter is a well-defined approximating functional consists of test functions such that $\varphi(0)=0$. It is clear that the only way to construct the extended approximating functional is by adding a $\delta$-function with an appropriate coefficient to the formal expansion. (This is similar to Bogoliubov's reasoning in his discovery of the correct form of the *R*-operation [1; 4]. In fact, Bogoliubov's construction served as a model for the theory of *As*-operation from the very beginning [39].) So, the complete expansion has the form:

$$\frac{1}{p^2-m^2} \approx \mathbf{As}\frac{1}{p^2-m^2}+O(m^4)=\frac{1}{p^2}+\frac{m^2}{p^4}+c(m)\delta(p)+O(m^4)\,. \qquad 3.5$$

where the coefficient $c(m)$ can be determined uniquely from the consistency conditions. In the present case the explicit expression in dimensionally regularized form is:

$$c(m)=\int \mathrm{d}p\frac{1}{p^2-m^2}\,. \qquad 3.6$$

Explicit expressions in a general Euclidean case can be found in [111] (the dimensionally regularized form) and [90; 102] (regularization-independent form). Here I'd like only to emphasize that the counterterms introduced by the *As*-operation serve two purposes: first, they contain divergent parts that kill the divergences of the formal expansion similarly to how the *R*-operation removes UV divergences in coordinate representation; second, the counterterms contain finite parts that are fixed uniquely (unlike the case of the *R*-operation) and their role is to fine-tune the expansion in order to ensure the correct approximation properties.

## Remarks 3.7

The above reformulation of the problem in terms of distributions is so important (the rest of the theory of *As*-operation is a deterministically logical development of this idea) that some philosophizing would be justified at this point:

(i) The problem 3.4 represents a generalization of the initial problem 3.1 of a kind when secondary but cumbersome details are ignored. As a result one is then able to concentrate on what is *really* important, whereby the problem gets simplified — even if unfamiliar notions become involved.

(ii) Our reformulation represents an embedding of the initial problem into a more general one. Such an embedding involves an arbitrariness which it is our right to fix as we please. In the present case, the arbitrariness involves choosing what test functions $\varphi(p)$ are allowed, whether we consider our linear functionals as continuous in any sense, etc. Again, such technicalities are less important than the heuristics of the general scenario and should be taken care of along the way so as to facilitate the complete solution. In our case, the most convenient choice [101] is to take the test functions and functionals from Schwartz' spaces $\mathcal{D}(P)$ and $\mathcal{D}'(P)$.

(iii) Dealing at first with test functions that vanish for all sufficiently large *p* corresponds to introducing a smooth UV cutoff. This agrees with the observation that whatever happens around $p=0$ has nothing to do with the behavior of the integral at large *p*. We will see below how the transition to integration over infinite regions is performed. Here it is important to note that introducing localized test functions allows one to concentrate on the vicinity of the singular point one wishes to study. This is just a more subtle implementation of the basic idea of the techniques of leading integration regions.

## Localization property and recursive structure of the *As*-operation 3.8

Consider a multiloop diagram whose denominator contains the same $m$-dependent propagator:

$$\int \mathrm{d}p \frac{1}{(p^2-m^2)} \int \mathrm{d}p' \frac{1}{(p-p')^2 \ldots}\,. \qquad 3.9$$

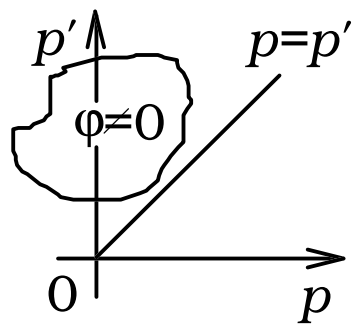

The singularities of the formal expansion of the $m$-dependent propagator are localized on the subspace $p = p'$ (the vertical axis). In accordance with the philosophy of the techniques of “leading integration regions”, consider contributions coming from vicinity of that subspace. From the point of view of distribution theory this means that one considers the expansion on test functions $\varphi(p, p' \ldots)$ that are zero near all other singularities of the integrand. Then the support of $\varphi$ (i.e. the region where $\varphi(p, p' \ldots) \neq 0$) can be as shown in the figure. It is clear that the product of such $\varphi(p, p' \ldots)$ and the rest of the integrand,

$$\varphi(p, p' \ldots) \times \frac{1}{(p-p')^2 \ldots}\,, \qquad 3.10$$

can be regarded as a smooth test function so that the problem degenerates to the case of a single propagator 3.4. Let us introduce the notation **As** for the operation that is applied to $m$-dependent products and returns their expansion in the sense of distributions. Then the above conclusions can be formally represented as:

$$\mathbf{As} \circ \left( \frac{1}{(p^2-m^2)} \times \frac{1}{(p-p')^2 \ldots} \right) = \left( \mathbf{As} \circ \frac{1}{(p^2-m^2)} \right) \times \left( \mathbf{T} \circ \frac{1}{(p-p')^2 \ldots} \right), \qquad 3.11$$

which is valid on test functions as described above. The operation **T** on the r.h.s. is the ordinary Taylor expansion in $m$. Its presence reflects the fact that the factors that expand without singularities in a given subregion do not require a special treatment.

More generally, let $G$ be a product of $m$-dependent propagators which one wishes to expand in $m$ in the sense of distributions, let $O$ be a subregion of the integration space where only some of the propagators from $G$ are singular (denote the product of such propagators as $G^{\mathrm{sing}}$), so that all other propagators from $G$ are regular (their product is denoted as $G^{\mathrm{reg}} \equiv G/G^{\mathrm{sing}}$). Then there emerges the following fundamental

| |
|---|
| Localization property of the *As*-operation: $\mathbf{As} \circ G\big\vert_O = \mathbf{T} \circ G^{\mathrm{reg}} \times \mathbf{As} \circ G^{\mathrm{sing}}$ |

3.12

This property explicates the fundamental recursion structure in our expansion problem. Its role in the expansion problem is similar to that of the microcausality condition in the construction of the $R$-operation [1; 58].

Recursive organization of a problem is clearly a benefit because it allows one to split the initial problem into similar but simpler subproblems, thus reducing the reasoning — whether at the heuristic stage or in the formal proof — to studying just one inductive step. All this is quite obvious and is exactly what the theory of *As*-operation is aimed at. Moreover, a paradigmatic example of such an approach was provided long ago by Bogoliubov’s construction of the ultraviolet $R$-operation [1; 58]. Recall in this respect the microcausality condition for the $S$-matrix which is equivalent to the following:

$$T[\mathcal{L}(x)\mathcal{L}(y)\mathcal{L}(z)\mathcal{L}(w)] = T[\mathcal{L}(x)\mathcal{L}(y)] \times T[\mathcal{L}(z)\mathcal{L}(w)], \quad x^0, y^0 > z^0, w^0\,. \qquad 3.13$$

In this relation, $T$ is the chronological product while $\mathcal{L}(x)$ is the density of the interaction Lagrangian at the space-time point $x$. If the recursive structure of the expansion problem becomes visible only after one adopts the point of view of the distribution theory, in the theory of $R$-operation it was on the surface all along: the original Bogoliubov's reasoning was clearly recursive and distribution-theoretic[i] even if it was never completely understood and appreciated (despite the obligatory ritualistic praises; note, incidentally, that ref. [1] is not quoted correctly even in the monograph [58]) — until used as an example to imitate in the expansion problem.[ii]

### Fundamental dilemma: singular factors *vs.* recursion 3.14

So, the fundamental underlying recursion (the localization property) involves *a product of singular factors*. Correspondingly, the dilemma is about how singularities are to be treated:

| | | |
|---|---|---|
| *Options:* | **BPHZ etc.**: sacrifice recursion to avoid distributions | ***As*-operation**: use distributions to preserve recursion |
| *Purpose:* | Formal rigor | Discoveries; calculational methods; meaningful proofs |
| *Pro:* | No fancy stuff! Absolutely convergent integrals only | Advantages of recursion fully used; **deterministic recipe** |
| *Contra:* | Extremely cumbersome; one has to know the answer; **not all recursions are resolvable** | Special techniques for handling singularities (a branch of distribution theory) |

3.15

### The conventional solution and the forest formula 3.16

For the reasons already discussed, it has long been taken for granted that all the expressions involving distributions should be reduced to ordinary absolutely convergent integrals by resolving the recursion. This approach was initiated by Bogoliubov and Parasiuk [4][iii]. It evolved into what is known as the BPHZ theory and became dominant for the following reasons.

First, the alternative approach requires a special technique that is not directly available in the standard toolkit of the distribution theory (cf. e.g. the inequalities that allow one to describe analytic structure of a singularity at an isolated point in presence of singularities localized on manifolds passing over that point [61; 101]). It is natural that people first tried to stay within the good old integral calculus even at the expense of heuristics and intuition.

The second reason was the general popularity of parametric representations. Recall that parametric representations were widely used in the 50s in studies of analytic properties of amplitudes (cf. e.g. the Landau equations [5]). The use of parametric representations seems to simplify the integrands (which is deceptive because the fundamental multiplicative structure is then lost). Also recall that the most successful methods of analytical calculations tend to avoid parametric representations — cf. e.g. the widely used algorithm of "integration by parts" [35][iv]). One the other hand, when one "studies parametric rep-

---

[i] One should distinguish Bogoliubov's heuristic derivation of the $R$-operation from how the latter is "proved" in the BPHZ theory.

[ii] *Note added*: Epstein and Glaser [15] did work out Bogoliubov's construction in a systematic mathematical fashion. Unfortunately, the distributional aspect was not discussed separately from other details such as operator definitions etc. As a result, the understanding achieved in [15] remained an esoteric knowledge. (I thank E. Kazes for pointing out to me ref.[15] and H. Epstein for a discussion.)

[iii] who, by the way, did not do that consistently with the resulting complication of the proof.

[iv] *Note added*: a far-reaching generalization is described in [118].

resentations” one is under a psychological illusion that something more meaningful is being done than developing a systematic notation for description of integrands and overcoming problems of iatrogenic character (the said destruction of multiplicative structure).

Lastly, Hepp [9] who corrected the proof of Bogoliubov and Parasiuk and introduced their result into the mainstream of Western theoretical tradition, was apparently not interested in distribution-theoretic aspect of the mechanism of *R*-operation and treated “Bogoliubov’s method” (i.e. the final prescription for the *R*-operation) rather formally as a “phenomenological procedure” to be “proved” (cf. the book [9], esp. Chapter 6 and the remarks preceding eq.(6.26)). As to Bogoliubov’s attitude, my guess would be that he probably needed, first and foremost, a perfunctory “rigorous proof” for a result whose correctness should have been clear from a combination of heuristic coordinate- and parametric-space arguments. Second, Bogoliubov apparently was — judging from the attention given to the $\alpha$-representation in the monograph [58] — fascinated, like many, by the interplay of combinatorial and graph-theoretic aspects of the $\alpha$-parametric representation.

Anyhow, the BPHZ tradition remained dominant for 30 years. The resulting multiple-summation formula which represents the solution of the relevant recursion in the Euclidean case is known as the forest formula [7; 71; 10].

It is true, of course, that there is no mystery about distributions, and general theorems of the distribution theory guarantee that any expression involving distributions can be rewritten in terms of ordinary integrals while the expansions reduce — at the most basic level — to subtractions, rearranging terms, splitting the integration domain into sectors à la Hepp, etc. But it is no less true, for instance, that any mathematical calculation reduces, in the final respect, to finite manipulations with integer numbers. No one would suggest, however, to forego the benefits of higher-level notions of rational numbers, calculus, etc., for the sole purpose of avoiding having to learn rules of rational arithmetic and infinitesimals.

On the other hand, not all recursions can be resolved into a comprehensible non-recursive formula (in our case, involving multiple summations over collections of subgraphs — forests). The Euclidean forest formula is already barely comprehensible.[i] In the non-Euclidean case the situation is much worse (as is clear already from the fact that the non-Euclidean case remains unsolved). The formal reason is as follows. Strictly speaking, the localization property is insufficient by itself. It allows one to determine only the structure of counterterms (cf. 3.5). Their coefficients (cf. 3.6) should be found using an additional reasoning based on the extension principle. The expressions for the coefficients close the recursion. In the Euclidean case such expressions are rather simple, especially within dimensional regularization (cf. 3.6). In the non-Euclidean case, expressions for the coefficients of counterterms involve secondary expansions so that the entire recursion is considerably more complex. Therefore, even if the non-Euclidean version of the forest formula could be obtained, I am highly skeptical about its usefulness even for the purposes of formal verification of results discovered by subtler methods.

This brings me to the next point. There is another specific reason why the forest-formula approach proved to be sterile as far as new results are concerned. A typical structure of a *R*- or *As*-operation (in distribution-theoretic notation) has the form

$$\mathbf{r}_G \circ \left( G \backslash \Gamma \sum_{\Gamma \subset G} \ldots \mathbf{r}_\Gamma \circ \Gamma \right), \qquad 3.17$$

where $\mathbf{r}_G$ is an operator which “takes care” of the singularity “associated” with a subproduct $G$.

---

[i] This, by the way, provides a psychological explanation of why partisans of the BPHZ theory insist so ferociously on formal rigor in studies of Feynman diagrams: When working with the forest formula one attempts to tackle the entire problem without splitting it into simpler subproblems; the underlying primitive power counting is thus obscured by the necessity to keep track of cancellations between seemingly amorphous collections of terms in the sum; as a result, the forest formula can only be dealt with in a meticulously formal manner. It is, therefore, not surprising that if the first thing one learns about multiloop diagrams is the forest formula, one then remains forever suspicious of approaches that promise better results with less pain.

Eq. 3.17 expresses the simple fact that before treating the singularity associated with the entire product $G$ (and localized, say, at an isolated point) one should take care of the singularities associated with subproducts $\Gamma \subset G$ (and localized on manifolds passing over that point; also note that Eq. 3.17 reflects the structure of the localization property). It should be emphasized that the explicit form of the operator $\mathbf{r}_G$ can only be determined after all the operators $\mathbf{r}_\Gamma$ for the subproducts have been constructed. However, the procedure of "resolving" the recursion involves expansion of the subtraction operators in presence of test functions, which can be represented in a symbolic form as follows:

$$\int (\mathbf{r} \circ \Gamma)\, \varphi \to \int \Gamma \times (\varphi - \mathbf{T}\varphi)\,, \qquad 3.18$$

where $\mathbf{T}$ is something like (not necessarily exactly) Taylor expansion. This means that in the resulting integrals *the order of subtractions is reversed* (in complete similarity with how the order of operators is reversed under conjugation). In the more general case of Eq. 3.17, the effect can be represented as follows:

$$\int X\, \mathbf{r}_G \circ \sum_{\Gamma \subset G} Y \mathbf{r}_\Gamma \circ Z \to \int \sum_{\Gamma \subset G} Z \times (1 - \mathbf{T}_\Gamma)[Y \times (1 - \mathbf{T}_G) X]\,. \qquad 3.19$$

(Note that the structure of the r.h.s. is clearly reminiscent of the forest formula; cf. the three-dot products of [58].) So, if one works with the resolved expressions, then the subtraction that has to be done first, $(1 - \mathbf{T}_G)$, corresponds to the singularity which should be analyzed last — after all other subtractions in subgraphs have already been determined.

> **The formal inversion of the order of subtractions** in subgraphs is the technical reason due to which the conventional methods do not allow one to make effective use of the recursion. This **explains their failure in the non-Euclidean case**. 3.20

On the other hand, within conventional approaches no attempts were ever made to study — and develop — the heuristics of the discoveries of the original results. One could recall in this respect that the heuristics which Bogoliubov used to find the correct formula for the $R$-operation was completely left out of the BPHZ theory. This is no wonder — while the proof given by Bogoliubov and Parasiuk was based on the first option of our dilemma, Bogoliubov's original reasoning [1] corresponds to the exactly opposite alternative.

## Solution of the asymptotic operation 4

Experts are well aware of the fact that any formal proof about multiloop diagrams reduces, in the final respect, to a primordial power counting. As I said, the conventional approach attempts to tackle the entire problem in an unstructured way so that power counting involves the entire expression with all the terms generated by all subtractions lumped together (e.g. the remainder of an expansion of a renormalized $l$-loop diagram).

> **The key analytical idea** of the theory of As-operation is to use the localization condition to structure the problem as an iteration of the same elementary step which involves **only a singularity localized at an isolated point**. 4.1

After this is understood, the problem practically reduces to identifying and analyzing a few representative few-loop examples (recall that a constructive distribution-theoretic recipe which is convenient in the case of an isolated singularity is given by the extension principle discussed above).

In the Euclidean case it is sufficient to consider two-loop diagrams[i]. The following two-loop example will help to illustrate this point. Consider the Euclidean diagram shown in Fig. 4.2. The two upper fat lines correspond to heavy particles of mass $M$ which is of order $Q$. The three other propagators contain light masses $m \ll Q, M$ . Let us route $Q$ through the fat lines. Then all the dependence on the heavy parameters will be conveniently localized in the two heavy propagators. The task is to expand the diagram in $m/M \to 0$ .

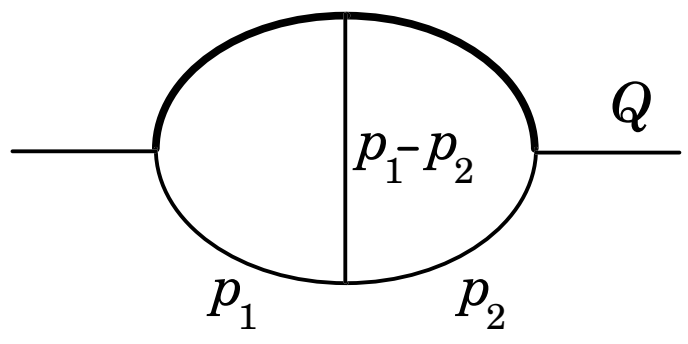

4.2

For simplicity I consider the two-dimensional case. Then there are no UV divergences, while nontrivial singularities emerge already in the leading $O(m^0)$ terms of the formal expansion in $m$. One can see that the two heavy propagators are independent of $m$ and constitute a "test function" as can be seen from the analytical expression:

$$\int d^2 p_1 \, d^2 p_2 \frac{1}{p_1^2 + m^2} \times \frac{1}{p_2^2 + m^2} \times \frac{1}{(p_1 - p_2)^2 + m^2} \times \frac{1}{(p_1 - Q) + M^2} \times \frac{1}{(p_2 - Q) + M^2} \ . \qquad 4.3$$

Quite obviously, the exact form of the "test function" is plays no role. So it is clear that it is sufficient to expand the product of the three $m$-dependent factors in the sense of distributions and insert the resulting expansion into the integral.

Construction of the expansion of the $m$-dependent factors in the sense of distributions is an iteration of the following sequence of steps:

- formal Taylor expansion;
- studying geometry of the singularities of the formal expansion;
- identifying collections of propagators (*IR*-subgraphs) whose singularities overlap at various points of the integration domain (*completeness condition*);
- studying analytical structure of the singularities (power counting);
- construction of counterterms.

*The formal expansion* has the form

$$\frac{1}{p_1^2 + m^2} \times \frac{1}{p_2^2 + m^2} \times \frac{1}{(p_1 - p_2)^2 + m^2} \approx \frac{1}{p_1^2} \times \frac{1}{p_2^2} \times \frac{1}{(p_1 - p_2)^2} + O\left(m^2\right) . \qquad 4.4$$

*The geometry of singularities* of the r.h.s. is shown in Fig. 4.5. Each denominator that can be equal to zero generates a manifold where the singularity is localized. In our case there are three different 2-dimensional subspaces that correspond to each of the three factors. Such singular manifolds can intersect. Each of the intersections should be counted as a singular manifold in its own right. This is because the nature of the singularity at the corresponding points can be non-trivial. "Non-

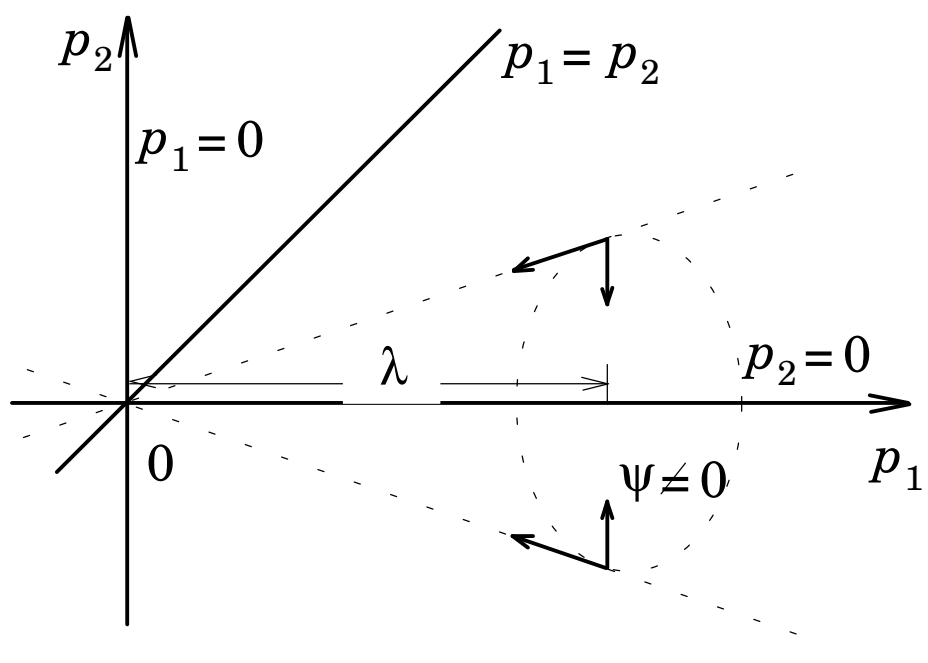

4.5

---

[i] Note that in order to see how the recursion will continue, it is necessary to consider certain properties of two-loop products that are not needed for expansions of two-loop integrals. This is still less cumbersome than considering the full three-loop case. More complicated examples do not exhibit new features. The non-Euclidean complications make already one-loop diagrams more difficult than Euclidean two-loop ones but our approach works in the non-Euclidean case, too.

trivial" means that the singularity does not factorize. In our case such an intersection is the point $p_1 = p_2 = 0$; the singularity is non-trivial because it cannot be factorized. In the absence of any one of the three factors this singularity becomes factorizable.

Enumeration of all intersections is easy: it is sufficient to enumerate all subsets of singular propagators subject to the restriction that momentum conservation imposes linear dependences on their momenta. This is known as *completeness condition* [55; 56; 111]. Indeed, if one puts to zero the momenta of the chosen collection of propagators (which corresponds to intersection of the corresponding manifolds) then there may be propagators whose momenta will automatically nullify owing to momentum conservation. This means that their singularities will overlap with the singularities of the collection and such propagators should be included in the collection. Collections of propagators that cannot be completed via the above procedure are exactly the *IR-subgraphs.* In the present case there are four *IR*-subgraphs: $\gamma_1$= (the propagator depending on $p_1$); $\gamma_2$= (the propagator depending on $p_2$); $\gamma_3$= (the propagator depending on $p_1 - p_2$); $\gamma_{123}$= (the collection of all three propagators). None of the three pairs of propagators constitute a complete collections and, therefore, an *IR*-subgraph.[i]

Next one should study the analytical nature of singularities. This is where the localization condition enters the game. Indeed, as explained above, the expansion valid on test functions $\psi$ that are non-zero only within the elliptical subregion in Fig. 4.5 is given by the formal expansion of the factors that are regular in the subregion, times the expansion in the sense of distributions of the propagator that is singular in the subregion. The expansion of an isolated propagator is known already (see 3.5; in the two-dimensional case the counterterm proportional to the $\delta$-function should be included already for the first term of the formal expansion which is logarithmically divergent at zero momentum). Therefore, the expansion in the subregion has the form (I am using the dimensional regularization)

$$\frac{1}{p_1^2}\times\left(\frac{1}{p_2^2}+c_0(m)\delta(p_2)\right)\times\frac{1}{(p_1-p_2)^2}+O\left(m^2\right)$$
$$=\frac{1}{p_1^2}\times\frac{1}{p_2^2}\times\frac{1}{(p_1-p_2)^2}+\frac{1}{p_1^2}\times c_0(m)\delta(p_2)\times\frac{1}{(p_1-p_2)^2}+O\left(m^2\right)\ . \qquad 4.6$$

The analysis of the vicinity of the singular manifold $p_2 = 0$ resulted in appearance of the counterterm on the r.h.s. It is easy to the consider in a similar manner the two singular manifolds $p_1 = 0$ and $p_1 = p_2$. One arrives at the expression

$$\frac{1}{p_1^2}\times\frac{1}{p_2^2}\times\frac{1}{(p_1-p_2)^2}+\frac{1}{p_1^2}\times c_0(m)\delta(p_2)\times\frac{1}{(p_1-p_2)^2}$$
$$+\frac{1}{p_1^2}\times\frac{1}{p_2^2}\times c_0(m)\delta(p_1-p_2)+c_0(m)\delta(p_1)\times\frac{1}{p_2^2}\times\frac{1}{(p_1-p_2)^2}+O\left(m^2\right)\ . \qquad 4.7$$

This is a correct expansion in the sense of distributions which is valid on test functions that are allowed to be non-zero everywhere except for a small neighborhood of the origin $p_1 = p_2 = 0$.

One can describe the effect of the above by saying that after the point $p_1 = p_2 = 0$ is removed, the three manifolds cease to intersect — the singularities "fall apart" — so that the resulting problem is effectively equivalent to simpler problems corresponding to singular manifolds of larger dimensionality and with lesser number of transverse dimensions, and with lesser number of singular factors contributing to that singularity.

It remains to transform the expansion 4.7 into an expansion that is valid everywhere in the integra-tion domain. This is done in two logical steps.

---

[i] If one uses a parametric representation then the analytical description of *IR*-subgraphs in terms of the completeness condition is insufficient. This is because the structure of integrands in parametric representations is characterized in a roundabout way via derivative graph-theoretic notions.

***First***, one notices that — similarly to the case of the *R*-operation in coordinate representation — the expansion 4.7 (*together* with the counterterms already added) is integrable with a wider set of test functions, not necessarily vanishing in an entire neighborhood of the origin. A proposition essentially equivalent to the Bogoliubov-Parasiuk theorem is that such less restrictive test functions should only have zero of second order (for a detailed discussion of the Bogoliubov-Parasiuk theorem in coordinate representation from the distribution-theoretic point of view see [61; 101]). It is natural to assume that the approximation properties of the expansion 4.7 are also preserved on such test functions. This is the crucial analytical step in the proof.

Assuming that this has been proved, the ***second*** step is then straightforward and consists in constructing the counterterm for the singularity at $p_1 = p_2 = 0$ in exactly the same way as in the case of the single propagator. Namely, the counterterm should be a linear combination of the $\delta$-function and its first-order derivatives while the coefficients have the form similar to 3.6. For instance, for the coefficient of the $\delta$-function without derivatives one has:

$$c_0^{\{\gamma_3\}}(m) = \int \mathrm{d}p_1\,\mathrm{d}p_2 \frac{1}{p_1^2+m^2} \times \frac{1}{p_2^2+m^2} \times \frac{1}{(p_1-p_2)^2+m^2} . \qquad 4.8$$

(For a discussion of such expressions see [111]).

Let us concentrate on the crucial analytical point of the above reasoning — power-counting in 4.7 at $p_1 = p_2 = 0$ in presence of the three counterterms. Consider again a test function localized within the elliptical region in Fig. 4.5. Introduce a parameter $\lambda$ as shown to parameterize the test function $\psi_\lambda(p_1, p_2) \equiv \psi(p_1/\lambda, p_2/\lambda)$ (*without* assuming anything about zeros of $\psi$). It is clear that studying the singularity of 4.7 at $p_1 = p_2 = 0$ is equivalent to studying the dependence of the value of the distribution 4.7 on $\psi_\lambda$ as $\lambda \to 0$. As already mentioned, a proposition essentially equivalent to the Bogoliubov-Parasiuk theorem is that — at fixed $m$ — the leading power behavior is $\lambda^{-2}$ i.e. the same as obtained by naive power-counting. On the other hand — at fixed $\lambda$ — the value of the remainder of the expansion (i.e. the difference between the initial expression, the l.h.s. of 4.4, and 4.7) is known to be $O(m^2)$. Now, what one needs to verify that the approximation property holds, is an estimate for the remainder combining *both* dependences in the factorized form, i.e. $O(\lambda^{-2}) \times O(m^2)$.

Note that such a property would mean that the type of the dependence on $\lambda$ and $m$ is *the same as in the first neglected term* — a property well-known in simpler cases of asymptotic expansions. Given that the analytic nature of the integrals we are dealing with — integrals of rational functions — is rather simple (despite the apparent combinatorial complexity), it is hard to imagine how it could have been otherwise. This may not be a “rigorous” — i.e. formalized — statement, but a heuristic argument like this is worth many a page of formal proofs. Anyhow, this type of estimates is certainly true for the individual propagators, and all one has to do is to perform a calculation that derives this property for the product as a whole. The recursion helps because one can use (here is the inductive assumption that will need to be verified) a factorized power-counting estimate for the single propagator (more, generally, the *IR*-subgraph corresponding to the singular manifold we are dealing with here).

It remains to note that a convenient formal language for precise description of such estimates and performing calculations of this type — essentially, power-counting in presence of logarithmic modifications plus the use of recursion — was developed in [61; 101].

Finally, a remark concerning combinatorics of how expansions in a global OPE-like form are restored. It is clear that the final formulas for expansions in the formalism of the *As*-operation run over *IR*-subgraphs which have a simple characterization (see above the completeness condition). Moreover, as can be seen from the expressions for counterterms, the latter have a direct interpretation as integrated Feynman diagrams corresponding to some matrix elements. This results in *drastic simplifications of combinatorics of exponentiation* as compared with the conventional reasoning (for details see [112]).

## Treatment of UV renormalized diagrams 4.9

Consider an unrenormalized Feynman diagram. Let $G(p, M_{tot})$ be its integrand where $p$ the collection of integration momenta and $M_{tot}$ the collection of all external momenta and masses of the diagram. Introduce a smooth cutoff at large $p$ ($\sim \Lambda$) in the form of a test function $\Phi(p/\Lambda)$ such that $\Phi(0) = 1$. Then letting $\Lambda \to \infty$ corresponds to removing the cutoff and restoring integration over the entire momentum space. Represent the cutoff as a sum over spherical layers:

$$\Phi(p/\Lambda) = \int_0^\Lambda \frac{d\lambda}{\lambda} \phi(p/\lambda) . \qquad 4.10$$

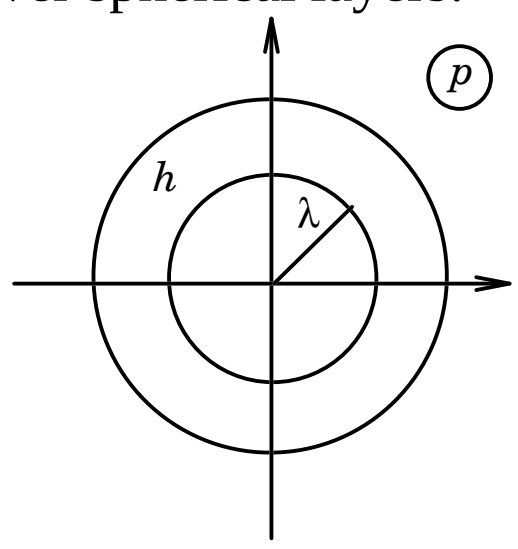

Changing the order of integrations over $p$ and $\lambda$ and performing rescaling $p \to \lambda p$, one can easily see that studying the behavior of the integral $\int dp\, \Phi(p/\Lambda) G(p, M_{tot})$ as $\Lambda \to \infty$ is equivalent to studying the expression $\int dp\, \phi(p) G(p, M_{tot}/\lambda)$ as $\lambda \to \infty$. This is exactly the same as studying the asymptotic expansion of the integrand of the latter in $M_{tot} \to 0$ in the sense of distributions over $p \neq 0$ (because $\phi$ are zero around $p = 0$). One can immediately use the formalism of the *As*-operation and obtain *a precise information about the UV behavior of the diagram*. It turns out that subtracting from the unrenormalized integrand all those and only those terms of the expansion that are responsible for UV divergences not only ensures UV convergence automatically, but also leads to a correct UV renormalization equivalent to the Bogoliubov *R*-operation [80; 91; 102].

One can show that studying expansion of renormalized diagrams then reduces to studying double *As*-operation with respect to a hierarchy of parameters (the light parameters of the expansion, and the totality of the dimensional parameters of the diagram, as shown above) [90; 102]. As above, all one has to do is to derive factorized estimates for the remainder of the double expansion. Such a problem is more cumbersome but presents no ideological problems.

## Theory of Euclidean asymptotic operation 4.11

A more or less complete theory of the *As*-operation has by now been developed for the Euclidean case, although the principles and much of the technique is completely general. The extension principle was found in [39; 45; 111]. The initial motivations were as follows:

- analysis of Bogoliubov's derivation of the *R*-operation [1; 58] in connection with multiloop calculations of the JINR and INR groups [27; 29; 34];
- a method of renormalization group calculations developed in [16; 24; 27; 34]; the method is based on the idea to exploit the simplicity of dependence of UV counterterms in the MS scheme on masses and momenta to simplify calculations. The method involved Taylor expansions in masses and momenta which sometimes led to difficulties due to IR divergences of the type discussed above;
- analysis of derivation of short-distance OPE with the aim of applying the calculational algorithms for 3-loop massless propagator-type integrals [35; 38; 43; 87] to problems involving operator product expansion; the standard formulas of the BPHZ theory proved to be of little use.

The motivations were of a distinctly applied nature. Correspondingly, the first results were devoted to practical calculations:

$R^*$ **-operation** [41] generalized the method of renormalization group calculations mentioned above. The heuristics behind the prescription of the $R^*$-operation are as follows. If one performs the expansion with all the appropriate counterterms as described above and takes into account the fact that the counterterms do not contain logarithmic dependence on dimensional parameters, then one can ignore the finite parts of the IR counterterms and choose them solely from condition of cancellation of IR divergences, ignoring the approximation property. Schematically:

$$\mathbf{R}_{\mathrm{MS}} \circ G \to \mathbf{As} \circ \mathbf{R}_{\mathrm{MS}} \circ G \to \tilde{\mathbf{R}} \circ \mathbf{R}_{\mathrm{MS}} \circ G \equiv \mathbf{R}^{*} \circ G \quad , \tag{4.12}$$

where $\tilde{\mathbf{R}}$ denotes the operation that differs from the *As*-operation by (absence of) finite parts. The resulting formulas allowed independent checks by direct calculations using different ways to expand in dimensional parameters. The definition of the *IR*-subgraph in [41] was not completely general because it resulted from a translation into the language of coordinate representation, which was motivated by an earlier calculational algorithm [33]. A general characterization was presented in [55; 56] (see the example of completeness condition given above as well as the discussion in [111]).

If $\alpha$-representation is used then the simple analytical characterization of IR subgraphs in terms of completeness condition is no longer sufficient. Therefore, a special graph-theoretic was invented [59]. Although of hardly any practical use because it is exceedingly cumbersome, it looks suitably obscure to convey an impression of rigor to a layman, and can be employed for the purposes of reference in papers claiming new proofs of the results discovered by other methods. The point here is that the integrand in $\alpha$-parametric representation is characterized indirectly using derivative graph-theoretic notions (cocycles etc.)[i]. Furthermore, after transition to parametric representation the simple multiplicative structure of momentum-space integrands is destroyed, and the structure of integrands is now described in terms of secondary graph-theoretic notions (two-trees, cocycles etc.). Correspondingly, the completeness condition which is simply and naturally formulated in the language of momentum representation [111] acquires an uncompromisingly rigorous aspect when rewritten in terms of cocycles [60].

From the analytical point of view, the essence of the mechanism of derivation of the method of *R**-operation reduces to commutativity of the double *As*-operation already discussed (complete proofs in a regularization-independent form as given in [102]).

There is a considerable body of calculations performed using the method of *R**-operation (see e.g. [51; 57; 84; 93; 94]). A calculation of a different kind based on a similar philosophy is presented in [85].

**Algorithms for OPE coefficient functions** [45; 50] within the MS scheme proved rather successful (see e.g. the recent 3-loop calculations [51; 57; 66; 86; 88; 89]). The theoretical and practical importance of the requirement of perfect factorization in OPE was first pointed out in [45; 44].

General formulas of asymptotic expansions for Euclidean asymptotic regimes, satisfying the criterion of perfect factorization and in a dimensionally regularized form most useful for applications, were presented in [56; 62; 63] (see also [112]) and discussed in the literature from the conventional point of view [68; 73; 78]. The combinatorics of exponentiation of expansions as worked out within the theory of *As*-operation (in particular, the method of inverse *R*-operation [67]) caused an overhaul of the treatment of combinatorial aspects within the BPHZ method as well [77].

A systematic description of the theory of Euclidean *As*-operation with emphasis on applications within dimensional regularization is presented in [111; 112].

Regularization-independent analysis of the entire theory was undertaken in [61; 69; 80; 82; 83; 90; 101; 102]. A new representation of UV renormalization has emerged within the framework of *As*-operation [69; 80; 102]. Such a representation gives a convenient way to formulate subtraction schemes such as the MS scheme [13] without using dimensional regularization. Developing purely 4-dimensional analytical formalisms is important in view of the fact that powerful computational methods based on the use of helicity amplitudes come to the forefront of perturbative QCD (see e.g. [70]).

[i] It is sometimes claimed that such an indirect (and extremely cumbersome) description is more “rigorous” than the simple analytical one. This is an example of how one sometimes confuses rigor with a meaningless abstruseness of unfortunately formalized “proofs”.

Apart from the papers [68; 73] already mentioned, there is a rather unusual series of papers [39; 45; 53; 52; 56; 62; 63; 111; 112] that were devoted to a verification of the formulae of Euclidean asymptotic expansions obtained within the framework of the theory of asymptotic operation. The papers were published by representatives of the Moscow school of $\alpha$-parametric representation (cf. ground-laying opus of that school [26]). The peculiarity of those papers is in that although they do not contain new results of any significance, they present formalized descriptions of the results already known — descriptions that are quite useful for the rigorous theory of $\alpha$-parametric representation.

An interesting mathematical problem of a general character is to analyze the use of dimensional regularization in applications directly in momentum space (cf. the original definition [11] and any application to perturbative QCD). The formal rewriting of results in the language of a parametric representation, as was done in [22; 77], remains highly unsatisfactory because it provides no insight whatsoever and is devoid of heuristic content. There are partial results in the right direction [42] but they remain insufficient.

## Towards non-Euclidean *As*-operation 4.13

The technical problems that are encountered in the extension of the method of the *As*-operation to non-Euclidean asymptotic regimes are due to the fact that the quadratic forms in denominators of Feynman propagators are no longer non-negative as in Euclidean case. Therefore, the singularities of individual factors are localized on second-order manifolds which can be singular in the differential-geometric sense (the apex of the light cone). The difficulties encountered are as follows:

**Osculating (="kissing") singularities** (non-generic intersections of singular manifolds). In order to analyze the nature of such singularities one has to perform straightening deformations of coordinates. Such deformations have to be found in an explicit form which is not always trivial. Another consequence is that some of the scalings that need to be done in order to perform power counting (cf. above) are asymmetric, i.e. the rate of scaling along different directions is different.

Possible problems due to **spurious algebraic dependences** between scalar invariants occurring in denominators. Recall that there always is a linear dependence between the vectors whose total number exceeds the dimension of the vector space. This might thwart any simple characterizations of *IR*-subgraphs. However, it looks like in many specific cases a complete analysis should be possible.

**Non-homogeneous expressions for counterterms** that result from the consistency conditions. Here a secondary expansion (so-called *homogenization*) needs to be performed. As a result one runs into factors that no longer correspond to standard propagators and exhibit an unusual UV behavior. Homogenization in a general case can be very tricky with somewhat unusual patterns of interplay between IR and UV divergences.

**Complex recursion** that does not exhibit factorization à la three-dot products of the BPHZ theory. This is probably an insurmountable obstacle for the conventional methods of proof.

I would like to conclude my talk with the following analogy which adequately illustrates some psychological aspects of how new paradigms are accepted:

| **Greek system/Latin notation** in Europe before IX c.: | | **Decimal system** brought from India by Arabs: |
|---|---|---|
| One can set to work right away **I+I=II** | | One has to learn first **1+1=2** |
| but | | but |
| things get tough soon: **MCLXXXII** | | then <u>*you can fly*</u> **1982** |
| *low-level languages* | *vs.* | *high-level languages* |
| **BPHZ** | | ***As*-operation** |

4.14

One might recall that in Florence even as late as 1299 AD, the use of decimal system in bank book-keeping was forbidden by law [6]. It is hard to believe now but decimal system was then considered complicated and confusing. Now we know that in the final respect, decimal system paved way to decimal fractions, binary system and floating point arithmetic built into any PC. Thinking is already hard enough for human beings; learning new ways to think is an order of magnitude harder (and mostly stops by 30). It also reminds me of some of the reports and citations that I received from BPHZ experts (see the quotation at the start of Sec. 3). The analogy also helps to understand e.g. why the common request to show a simple one-loop example cannot be satisfied easily — this is because analysis of simple one-loop examples consists mostly of definitions whose role becomes clear only in the analysis of higher-order corrections, while the simpler one-loop examples are perfectly clear without any distribution theory. (What is the advantage of using 1+1=2 instead of I+I=II? )

Nevertheless, as of the present moment, most of the problems that I enumerated in the discussion of the non-Euclidean case, seem to have concrete solutions. I hope to be able to talk about them in a not too distant future.

(*Note added*. The non-Euclidean *As*-operation has been successfully constructed in F.V.Tkachov, Phys. Lett. B412 (1997) 350 at the level of generality that could not have been envisaged at the time of preparation of this review.)

## References 5